\documentclass[amsmath,amssymb,12pt,superscriptaddress,nofootinbib]{revtex4-1}
\usepackage{amsmath,amssymb}
\usepackage{graphicx}
\usepackage{subfigure}
\usepackage{color}
\usepackage{ulem}
\renewcommand\b{\begin{equation}}
\newcommand\e{\end{equation}}
 
\newcommand\n{\nonumber}
\renewcommand\([1]{\left(#1\right)}
\renewcommand\[{\left[}
\renewcommand\]{\right]}
\newcommand{\dd}{\mathrm{d}}

\newcommand{\ee}{{\rm e}}
\definecolor{DarkBlue}{rgb}{0,0,0.7}

\definecolor{DarkRed}{rgb}{0.65,0,0} 

\begin{document}
\baselineskip5.5mm
\thispagestyle{empty}

{\baselineskip0pt
\small
\leftline{\baselineskip16pt\sl\vbox to0pt{
               \hbox{\it Division of Particle and Astrophysical Science, Nagoya University}
               \hbox{\it Graduate School of Sciences and Engineering for innovation, Yamaguchi University}
%
                             \vss}}
\rightline{\baselineskip16pt\rm\vbox to20pt{
            {
            }
\vss}}%
}

\author{Tomohiro~Nakamura}\email{nakamura.tomohiro@g.mbox.nagoya-u.ac.jp}
\affiliation{
Division of Particle and Astrophysical Science,
Graduate School of Science, Nagoya University, 
Nagoya 464-8602, Japan
}

\author{Taishi~Ikeda}\email{ikeda@gravity.phys.nagoya-u.ac.jp}
\affiliation{ 
Division of Particle and Astrophysical Science,
Graduate School of Science, Nagoya University, 
Nagoya 464-8602, Japan
}

\author{Ryo~Saito}\email{rsaito@yamaguchi-u.ac.jp}
\affiliation{
Graduate School of Sciences and Technology for Innovation, 
Yamaguchi University, Yamaguchi 753-8512, Japan
}

\author{Chul-Moon~Yoo}\email{yoo@gravity.phys.nagoya-u.ac.jp}
\affiliation{
Division of Particle and Astrophysical Science,
Graduate School of Science, Nagoya University, 
Nagoya 464-8602, Japan
}

\vskip1cm

\title{Chameleon Field in a Spherical Shell System}
\begin{abstract}


We study the  
screening mechanism 
of a chameleon field in a highly inhomogeneous density profile.
For simplicity, we consider 
static and spherically symmetric systems 
which 
are composed 
of concentric  
thin shells. 
We calculate the fifth force profile with different methods depending on the the Compton wavelength of 
the chameleon field:
a numerical method for relatively large values of the Compton wavelength and 
an analytic approximation for the small Compton wavelength limit. 
Our results show that, if the thin-shell condition for the corresponding smoothed density profile 
is satisfied, the fifth force is safely screened outside the system 
irrespective of the configuration of the shells inside the system. 
In contrast to the 
outer region,
we find that the fifth force can be comparable to the Newtonian gravitational force 
 in the interior region. 
This is simply because each shell is unscreened in thin shell limit 
even though the density of the shell is infinitely large. 
Our results explicitly show that
the screening mechanism successfully works for a cluster of unscreened objects 
if the cluster itself satisfies the thin-shell condition on average. 
At the same time, 
even when the screening mechanism is working for a total system, 
its components can be unscreened and then a large fifth force can appear in its inside. 
One should not feel complacent about the wellbehavedness of the fifth force field 
with an averaged density distribution when we consider highly inhomogeneous system. 
\end{abstract}
\maketitle


\section{Introduction}
\label{intro}
General relativity
is the most successful theory of gravity, which can explain various gravitational 
phenomena including gravitational waves recently observed for the first time \cite{GW1,GW2}.
There is no 
experimental 
result which clearly contradicts
general relativity so far. 
Nevertheless, 
many people are fascinated by the fundamental question of how accurately general relativity describes our universe, 
and attracted by seeking for an alternative gravitational theory 
as a clue of quantum gravity or to elucidate the dark side of our universe.

Scalar-tensor theories of gravity are simple examples of the modified gravity, 
which were originally proposed by Brans and Dicke in 1961~\cite{Brans1961}. 
It contains an additional scalar field other than the Einstein-Hilbert term and the standard matter term. 
One of the motivations to consider scalar-tensor theories is 
to explain the accelerated expansion of the universe by adding the 
scalar degree of freedom. 
The scalar field usually couples to the standard model particles and 
affects the motion of them
through the so-called fifth force.
Experimental tests of gravity in the solar system can give strong constraints on the fifth force 
and thus parameters of scalar-tensor theories \cite{Will2014}. 

In order to accord with the fifth force constraints, 
scalar-tensor theories must have a mechanism that screens the fifth force mediated by the scalar field on small scales.
We may classify the scalar fields by mechanisms of the screening~\cite{Joyce2015}.
One example is the chameleon field introduced by 
Khoury and Weltman~%
\cite{Khoury2004a,Khoury2004}. 
The chameleon field has a large 
value of the effective mass in a sufficiently high density region such as on the Earth 
or in the solar system,
so that the fifth force mediated by the chameleon field becomes an unobservable short-range force. 
In contrast, the chameleon field has a
smaller mass and long Compton wavelength in cosmological low density regions so that it could accelerate 
the expansion of our universe. 
This chameleon mechanism can be also applied to other types of modified gravity theories 
such as $f(R)$ gravity (see, e.g. Ref. \cite{Felice2010}). 
A lot of experimental tests have been 
proposed and performed in order to seek such a field, e.g. astrophysical tests 
such as those using distance indicators~\cite{Jain2013} or Galaxy rotation curve~\cite{Vikram2014}, 
and laboratory tests such as those using torsion pendulum~\cite{Upadhye2012}, atom interferometer~\cite{Hamilton2015}, and so on.

Calculations of the fifth force have been mainly done with 
a spherically symmetric smooth density profile surrounded by a cosmological low density region 
as the environment. 
For instance, for a compact object,  
we can estimate the scalar charge and show the fifth force can be much  
weaker than the Newtonian gravitational force (see, e.g. Ref. \cite{Khoury2004} and Refs. \cite{Kobayashi2008, Upadhye2009, Tsujikawa:2009yf, Babichev:2009td, Babichev:2009fi} for relativistic stars).
Recently, 
the chameleon mechanism in more general situations has been started to be investigated in numerical ways. 
The screening effect on the structure formation is investigated by generalized N-body simulations \cite{Oyaizu2008f, Oyaizu2008s, Oyaizu2008t, Zhao2011, Zhao2012}
and
strong constraints on $f(R)$ parameters
are obtained from the modified gravity effects on galaxy clusters \cite{Terukina2014,Wilcox2016}. 
Also, the screening for non-spherical sources is investigated in Ref. \cite{Burrage2018}.
In this paper, we focus on an aspect that has been overlooked in the above analyses. 
Usually, the screening effect for a system is investigated by using the smoothly averaged density profile
over the system. 
However, actual objects in the universe 
do not necessarily have a smooth density profile but inhomogeneous in general. 
If the Compton wavelength of the field is shorter than scales of the inhomogeneities,   
the smoothing may not be justified and 
effects of the inhomogeneities should be taken into account.   
For example, in our galaxy, the upper bound on the Compton wavelength of the chameleon field can be obtained as $\lambda_{\phi} < 10^{7-12} {\rm m}$ \cite{Davis2014} by rescaling the terrestrial experimental upper bound, 
which is smaller than the average interstellar distance. 
Also in globular clusters, 
the Compton wavelength $\sim 10^4 {\rm m}$, is much less than distances between the stars in the cluster.
This indicates that the chameleon field may vary
rapidly and be kicked by the inhomogeneity. 
Then, a significant fifth force may be mediated inside an inhomogeneous object like a galaxy.

In order to 
understand the essence of effects of inhomogeneities, as a first step, in this paper, 
we keep the system as simple as possible with extremely large density contrasts. 
Concretely, we assume a static spherically symmetric system composed of a 
set of infinitely thin shells at regular intervals of radius,
where the inhomogeneity is controlled by the number of the shells. 
The shell interval corresponds to the scale of the inhomogeneity in this system.
Thus, if we choose the parameters such that the Compton wavelength is shorter than the
shell interval, the scalar field is perceptible to the inhomogeneity and a 
significantly large fifth force may appear inside the system.  
Moreover, the fluctuations of the field inside the system may also affect the scalar charge of the overall system. 
We calculate the field profile and the fifth force strength,
and investigate those dependence on the parameters of the system.

This paper is organized as follows. 
In the section \ref{chamfield}, we introduce the chameleon field and the fifth force. 
A brief review of the 
uniform density case is given in the section \ref{constar} for comparison with our case. 
Then, we introduce our model, the spherical shell system in the section \ref{SSS}.
The resultant fifth-force profiles are 
shown in the section \ref{SinSSS}.
In the section \ref{thick}, 
we investigate how the fifth-force profiles change 
as the shells become thicker.
Section \ref{conclusion} is devoted to a summary and conclusion. 
In this paper, we use natural units in which both
the speed of light $c$ and the reduced Planck constant $\hbar$ are 
one.


\section{Chameleon Field}
\label{chamfield}

A prototype of the chameleon field is given by a scalar field with a conformal coupling and a runaway-type potential
\cite{Khoury2004},
\b
\nabla_{\mu}\nabla^{\mu}\phi-\frac{\beta}{M_{\rm pl}}\rho - V'(\phi) =0 \,,
\e
where $\beta$ represents a dimensionless conformal coupling and the potential $V(\phi)$ is typically assumed to be 
the inverse power-law potential: $V(\phi) = M^{4+n}/\phi^{n}$.
The prime means the derivative with respect to $\phi$. 
Here, $M_{\rm pl}$ is the Planck mass and $M$ is bounded above as $M \lesssim 10^{-3} {\rm eV}$ 
to evade laboratory constraints on the fifth force \cite{Khoury2004}, 
where $\beta$ is assumed as ${\cal O} (1)$.  
The second and the third terms can be combined into derivative of 
the following effective potential:
\b \label{def:potential}
V_{\rm eff}(\phi) \equiv \frac{\beta}{M_{\rm pl}}\rho\ \phi+\frac{M^{4+n}}{\phi^n}.
\e
This effective potential has the minimum at 
\b \label{def:minimum}
\phi_{\rm min}(\rho) \equiv M \({\frac{nM^3 M_{\rm pl}}{\beta\rho}}^{\frac{1}{n+1}},
\e
and the mass around this minimum is evaluated as 
\begin{eqnarray}
	m_{\rm eff}^2(\rho) &\equiv& V''_{\rm eff}(\phi_{\rm min}) \nonumber \\
	&=& \frac{(n+1)\beta \rho}{M M_{\rm pl}} \left( \frac{\beta\rho}{nM^3 M_{\rm pl}} \right)^{\frac{1}{n+1}} \,. \label{def:mass}
\end{eqnarray}
The effective mass of the chameleon field increases with $\rho$. 

In the static and spherically symmetric case, 
the equation of motion (EoM) becomes 
\b
\frac{\dd^2\phi}{\dd r^2}+\frac{2}{r}\frac{\dd\phi}{\dd r}-\rho\frac{\beta}{M_{\rm pl}}+n\frac{M^{4+n}}{\phi^{n+1}}=0.
\e
For later convenience, 
we rewrite the above equation by 
the following 
dimensionless variable:
\b
\hat{\phi}\equiv\frac{\phi}{\phi_{\rm c}},
\e
where $\phi_{\rm c}$ is the field value at the potential minimum (\ref{def:minimum}) for the central density $\rho_c$, 
that is, 
$\phi_{\rm c} \equiv \phi_{\rm min}(\rho_{\rm c})$.
In addition, we introduce a length scale $L$ and use the normalized radius $x$ defined by $x \equiv r/L$.
Then, we obtain 
\begin{eqnarray}
\frac{\dd^2\hat{\phi}}{\dd x^2}+\frac{2}{x}\frac{\dd\hat{\phi}}{\dd x}-\hat{\rho}\tilde{m}_{\rm c}^2 L^2 +\frac{\tilde{m}_{\rm c}^2 L^2}{\hat{\phi}^{n+1}}&=&0,
\label{chameq}
\end{eqnarray}
where $\tilde{m}_{\rm c}^2\equiv \beta\rho_{\rm c}/M_{\rm pl}\phi_{\rm c}$ and $\hat{\rho}\equiv\rho/\rho_{\rm c}$. 
Note that 
$m_{\rm eff}^2(\rho_{\rm c}) = (n+1)\tilde{m}_{\rm c}^2$.

As will be reviewed in the next section, 
the large effective mass (\ref{def:mass}) can enforce the chameleon field to be approximately fixed at the minimum $\phi_{\rm c}$ in the interior of the star 
and the matter inside the star does not contribute to the scalar charge except for the thin outer shell region 
whose width is comparable to the Compton wavelength.
Therefore, the fifth force 
\b
F_{\phi}=\frac{\beta}{M_{\rm pl}}\frac{\dd\phi}{\dd r} 
=\frac{\beta\phi_{\rm c}}{M_{\rm pl}}\frac{\dd\hat{\phi}}{\dd r} \,,
\label{fforce}
\e  
is screened by the $\rho$-dependent mass.
However, this argument is based on the smoothed density. 
When density contrasts are high, the potential minimum (\ref{def:minimum}) and the effective mass (\ref{def:mass}) will vary rapidly.
For such a system, it will not be appropriate to solve the EoM with the smoothed density 
and the inhomogeneity should be taken into account.
Unlike the smooth density case, the field value  
may vary also in the interior of the system, which causes the appearance of a significant fifth force.


\section{Field Profile of Constant Density Stars}
\label{constar}
We review how the chameleon field is sourced by a star with the constant density $\rho_{\rm c}$ 
surrounded by the cosmological density $\rho_{\infty}$. 
We take the radius of the star as the unit of the length scale $L$. 
If the effective mass \eqref{def:mass} for the constant density $\rho_{\rm c}$ is sufficiently large, 
the field value stays near the potential minimum $\phi_{\rm c}$ around the center of the star. 
Then, we assume that there is a radius from which the field value starts to change and denote 
this radius as $x_{\rm roll}$.
We can divide the whole region into 
the following three pieces.

\begin{enumerate}

\item$x<x_{\rm roll}$\\
In this region, the value of the chameleon field does not change much, 
and its value $\hat{\phi}$ and the first derivative $\dd\hat{\phi}/{\dd x}$ can be approximated by one and zero, respectively. 

\item$x_{\rm roll}<x<1\ (r_{\rm roll}<r<L)$\\
The chameleon field rolls down the effective potential toward a larger value. 
Then, the first term in the effective potential is dominant, so that  
the EoM becomes
\b
\frac{\dd^2\hat{\phi}}{\dd x^2}+\frac{2}{x}\frac{\dd\hat{\phi}}{\dd x}=\tilde{m}_{\rm c}^2L^2.
\label{bound}
\e
The solution of the equation~(\ref{bound}) with the boundary conditions $\hat{\phi}=1$ and $\dd\hat{\phi}/\dd x=0$ at $x=x_{\rm roll}$ is given by
\b
\hat{\phi}=1+\frac{\tilde{m}_{\rm c}^2L^2}{6}\({\frac{2x_{\rm roll}^3}{x}+x^2-3x_{\rm roll}^2}.
\label{phic}
\e

\item$x>1\ (r>L)$\\
The chameleon field  
quickly 
falls into the value sufficiently 
close to the minimum for the cosmological background $\hat{\phi}_{\infty}\equiv(\rho_{\rm c}/\rho_{\infty})^{\frac{1}{n+1}}$ outside the star. 
Then, an approximate solution is obtained by linearizing the EoM (\ref{chameq})
and we obtain 
\b
\hat{\phi}=\hat{\phi}_{\infty}+A\frac{\ee^{-m_{\infty}Lx}}{x},
\label{phio}
\e
where $m^2_{\infty}\equiv m_{\rm eff}^2(\rho_{\infty})$.
\end{enumerate}
Matching $\hat{\phi}$ and $\dd\hat{\phi}/\dd x$ at $x=1$ by using the equations (\ref{phic}) and (\ref{phio}), we  obtain 
\b
A=-\frac{\tilde{m}_{\rm c}^2L^2}{3}\frac{1-x_{\rm roll}^3}{m_{\infty}L+1}\ee^{m_{\infty}L} \,,
\label{A}
\e
and
\b
\hat{\phi}_{\infty}-1=\frac{\tilde{m}_{\rm c}^2L^2}{6}\({2\frac{1-x_{\rm roll}^3}{m_{\infty}L+1}+1+2x_{\rm roll}^3-3x_{\rm roll}^2}.
\label{phiinf}
\e
If the density of the object is sufficiently large, it is expected that the chameleon field stays near the minimum $\hat{\phi} \simeq 1$ in almost whole region inside the star,
and then $x_{\rm roll} \simeq 1$.
This limit is so-called
the thin shell regime because the only thin shell part of the star ($x_{\rm roll}<x<1$) contributes to the exterior field profile. 
Then, the equations~(\ref{A}) and (\ref{phiinf}) can be approximated as follows:
\b
A\simeq-\tilde{m}_{\rm c}^2L^2\frac{1-x_{\rm roll}}{m_{\infty}L+1}\ee^{m_{\infty}L}, 
\label{eq:A}
\e
\b
1-x_{\rm roll}\simeq\frac{1}{\tilde{m}_{\rm c}^2L^2}(m_{\infty}L+1)(\hat{\phi}_{\infty}-1).
\label{thinpara}
\e
We obtain the approximate form of $\hat{\phi}$ by substituting  
equations \eqref{eq:A} and \eqref{thinpara} into the equation (\ref{phio})
as follows:
\b
\hat{\phi}=\hat{\phi}_{\infty}-(\hat{\phi}_{\infty}-1)\frac{\ee^{-m_{\infty}L(x-1)}}{x}.
\label{ffcon}
\e
We can check that the equation (\ref{thinpara}) is consistent 
with the assumption $x_{\rm roll} \simeq 1$ if the following condition is satisfied:
\b
\frac{\hat{\phi}_{\infty}-1}{\tilde{m}_{\rm c}^2L^2}(m_{\infty}L+1)\ll1 \,,
\label{shelcon}
\e
which is satisfied when the Compton wavelength $\lambda_\phi \equiv 1/m_{\rm c}~(m_{\rm c} \equiv \sqrt{n+1}\tilde{m}_{\rm c})$ is much shorter than the radius of the star $L$.

From the equation (\ref{fforce}), the fifth force for the constant density star is calculated as,
\b
F_{\phi}=F^{\rm con}_{\phi}:=\frac{\beta\phi_{\rm c}}{M_{pl}L}(\hat{\phi}_{\infty}-1)(m_{\infty}L x+1)\frac{\ee^{-m_{\infty}L(x-1)}}{x^2}.
\e
The chameleon field has a sufficiently small effective mass (\ref{def:mass}) in the cosmological background unless $M$ is too small.
Then, the Compton wavelength of the chameleon field in the cosmological background is much longer than the radius of the star $L$, e.g. $\lambda_\phi \sim {\rm 1 Mpc}$ for $M \sim 10^{-3} {\rm eV}$.
Taking the limit $m_\infty L\rightarrow 0$, 
we obtain the following expression: 
\begin{equation}
\lim_{m_\infty L\rightarrow 0}F^{\rm con}_{\phi}=\frac{\beta\phi_{\rm c}}{M_{\rm pl}L}(\hat{\phi}_{\infty}-1)\frac{1}{x^2}.
\label{fiff_uni}
\end{equation}
Since the Newtonian gravitational force made by the constant density star is 
given by 
\b
F_{\rm Newton}=\frac{1}{8\pi M_{\rm pl}^2}\frac{4\pi L^3\rho_{\rm c}}{3L^2x^2}\n
=\frac{\rho_{\rm c}L}{6M_{\rm pl}^2}\frac{1}{x^2}, 
\e
we can evaluate the ratio $\mathcal R$ between the fifth force and the Newtonian gravitational force, which corresponds to the scalar charge in units of the stellar mass, as
\begin{eqnarray}
\mathcal R:=\left|\frac{F_{\phi}}{F_{\rm Newton}}\right|&=&6\frac{\beta M_{\rm pl}\phi_{\rm c}}{\rho_{\rm c}L^2}(\hat{\phi}_{\infty}-1)(m_{\infty}L x+1)\ee^{-m_{\infty}L(x-1)}\n\\
&\simeq&6\beta^2\frac{\phi_{\rm c}}{\tilde{m}_{\rm c}^2L^2}(\hat\phi_{\infty}-1),
\label{thin}
\end{eqnarray}
where we have taken the limit $m_\infty L\rightarrow 0$ in the second line.
We can find that, if the thin-shell assumption is valid, that is, 
the equation \eqref{shelcon} is satisfied, 
the value of $\left|F_{\phi}/F_{\rm Newton}\right|$ is suppressed.

On the other hand, in the case $x_{\rm roll}\simeq 0$, which is called the thick-shell limit, 
the field value 
and the fifth force are given by, respectively, 
\b
\hat{\phi}=\hat{\phi}_{\infty}-\frac{\tilde{m}_{\rm c}^2L^2}{3(m_{\infty}L+1)}\frac{\ee^{-m_{\infty}L(x-1)}}{x},
\e
\b
\mathcal R=2\beta^2\frac{m_{\infty}Lx+1}{m_{\infty}L+1}\ee^{-m_{\infty}L(x-1)} = {\cal O}(\beta^2), 
\e
where we have assumed $m_{\infty}L\simeq0$.


\section{Spherical Shell System}
\label{SSS}

\begin{figure}[t]
\begin{center}
\includegraphics[clip,width=8cm]{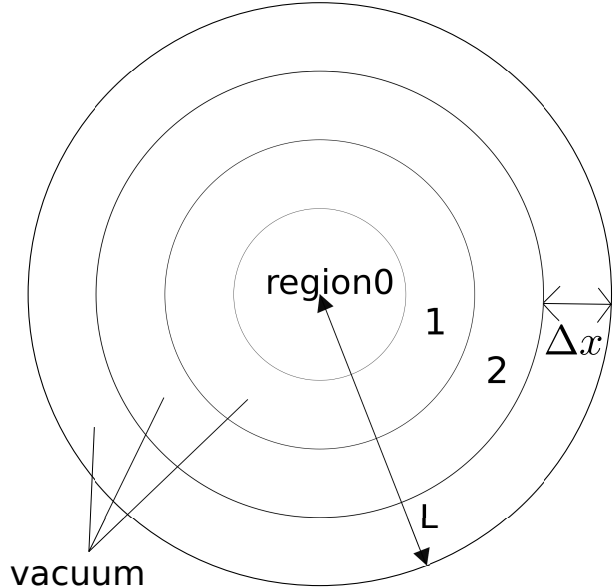}
\caption{A schematic figure of the spherical shell system.}
\label{sss}
\end{center}
\end{figure}

In the previous section, 
we have considered a spherical object with a constant density, where 
the large effective mass can make the chameleon field stay at the minimum of the effective potential $V_{\rm eff}$ inside the object. 
In this section, we consider a simple but non-trivial example of an inhomogeneous system:
$N$ pieces of concentric spherical shells 
separated by vacuum regions with regular intervals $\Delta x$ and equal surface density $\sigma$ (see Fig.~\ref{sss}). 
The shells are assumed to be infinitely thin, that is, the radial density profile of each shell is approximated by a delta function. 
Moreover, in order to avoid running away of the chameleon field to infinity,  
we assume that the shell system is surrounded by the cosmological density $\rho=\rho_{\infty}$ as usual. 
Under these idealizations, we investigate how the inhomogeneities can have an impact on the field profile.
In this system, 
neither the potential minimum nor the effective mass is defined at any radius and the previous intuitive argument cannot be applied.
In reality, we would
need to introduce a small density between the shells.
Nevertheless, if the field value does not reach the minimum of the effective potential in the intervals, 
it is irrelevant whether the density is finite or zero as we assumed.
In addition, the infinitely thin shells are an idealization. 
We will discuss how our argument here is affected when the shells are thicker in the section~\ref{thick}.

We assumed that 
the shells regularly foliate the spherical region with a fixed interval of the radius,
and the surface density of each shell is identical to each other. 
Then, given the radius of the outermost shell $L$, the interval $\Delta x$ can be written as $L/N$. 
Denoting the total mass of the system by $M_{\rm tot}$, 
the surface density $\sigma$ is given by,
\begin{eqnarray}
\sigma&=&\frac{M_{\rm tot}}{4\pi (L/N)^2(1^2+2^2+\cdots+N^2)}\n\\
&=&\frac{3M_{\rm tot}N}{2\pi L^2(N+1)(2N+1)} \,.
\end{eqnarray}
The smoothed density $\rho_{\rm c}$ can be written as 
$\rho_{\rm c}=3M_{\rm tot}/4\pi L^3$ and is related to the surface density as
\b
\sigma = \rho_{\rm c}L\frac{N}{(N+1)(N+1/2)} \,.
\e
Hereafter, we use the radius of the outermost shell $L$ as the length scale $L$ in the section {\ref{chamfield}}.

Under these setup, the field equation (\ref{chameq}) becomes
\b
\frac{\dd^2\hat{\phi}}{\dd x^2}+\frac{2}{x}\frac{\dd\hat{\phi}}{\dd x}+\frac{\tilde{m}_{\rm c}^2L^2}{\hat{\phi}^{n+1}}=0, 
\label{vacuum}
\e
in the vacuum regions,
and 
\b
\frac{\dd^2\hat{\phi}}{\dd x^2}+\frac{2}{x}\frac{\dd\hat{\phi}}{
\dd x}+\frac{\tilde{m}_{\rm c}^2L^2}{\hat{\phi}^{n+1}}-\hat{\rho}_{\infty}\tilde{m}_{\rm c}^2L^2=0 \,,
\label{outside}
\e
in the outer cosmological region $r > L$.
The junction condition at each shell is given by \cite{Deruelle2008}
\b
[\phi]^+_-=0 \,,
\label{bfp}
\e
\b
\[\frac{\dd\phi}{\dd r}\]^+_-=\beta\frac{\sigma}{M_{\rm pl}}.
\e
The symbol $[~]^+_-$ on the left hand side of the equations is defined by
\b
[f(x)]^+_-\equiv \lim_{x\rightarrow x_{\rm shell}+0}f(x)-\lim_{x\rightarrow x_{\rm shell}-0}f(x).
\e
It will be more suggestive to rewrite the surface density in the second junction condition in terms of the smoothed density $\rho_{\rm c}$ or the effective mass $\tilde{m}_{\rm c} ~ (\equiv m_{\rm c}/\sqrt{n+1})$:
\begin{eqnarray}
\[\frac{\dd\hat{\phi}}{\dd x}\]^+_-&=&\beta\frac{\rho_{\rm c}L^2}{M_{\rm pl}\phi_{\rm c}}\frac{N}{(N+1)(N+1/2)}\n\\
&=&\tilde{m}_{\rm c}^2L^2\frac{N}{(N+1)(N+1/2)} \,. \label{bfdp}
\end{eqnarray}

The Newtonian gravitational force in the $i$-th region 
is given by
\begin{eqnarray}
F_{\rm Newton}&=&\frac{1}{8\pi M^2_{\rm pl}}M_{\rm tot}\frac{i(i+1)(2i+1)}{N(N+1)(2N+1)}\frac{1}{r^2}\n\\
&=&\frac{\rho_{\rm c}L}{6M^2_{\rm pl}}\frac{i(i+1)(2i+1)}{N(N+1)(2N+1)}\frac{1}{x^2}\hspace{2cm}
{\rm for} \quad i/N<x<(i+1)/N,
\label{newton}
\end{eqnarray}
where
$i$ 
runs over $0$ to $N$. 
As is well known, the Newtonian gravitational force $F_{\rm Newton}$ depends only on the enclosed mass at a given radius irrespective of its internal structures.
Then, from the equation (\ref{fforce}), we obtain 
\begin{eqnarray}
\mathcal R
&=&6\beta\frac{M_{\rm pl}\phi_{\rm c}}{\rho_{\rm c}L^2}\frac{N(N+1)(2N+1)}{i(i+1)(2i+1)}\frac{\dd \hat{\phi}}{\dd x}x^2\n\\
&=&\frac{6\beta^2}{\tilde{m}_{\rm c}^2L^2}\frac{N(N+1)(2N+1)}{i(i+1)(2i+1)}\frac{\dd \hat{\phi}}{\dd x}x^2.
\end{eqnarray}

To see the impact of the inhomogeneities, in the next section, we will evaluate the value of $\dd\phi/\dd x$ 
for various values of the parameters $\tilde{m}_{\rm c}L$ and $N$, 
which represent the ratio of the length scales in the system, $L/\lambda_\phi$ and $L/\Delta x$, respectively. 
For the smoothed density, 
we found $x^2 \dd \hat{\phi}/\dd x \sim \hat{\phi_\infty} - 1$,
and thus the small factor $1/\tilde{m}_{\rm c}^2L^2 \simeq (\lambda_\phi/L)^2$ ensures the screening.
On the other hand, in our case, 
when the interval $\Delta x$, is large enough,  
the chameleon field is expected to vary rapidly and the ratio ${\cal R}$ might become large.


\section{Screening in the Spherical Shell System}
\label{SinSSS}

We solve the field equations (\ref{vacuum}) 
and (\ref{outside}) with the junction conditions (\ref{bfp}) and (\ref{bfdp}) 
taking into account the thin-shell condition (\ref{shelcon}) for the smoothed density. 
Here, as an example,  
we consider 
the averaged density of a galaxy for $\rho_{\rm c}$ as 
$\rho_{\rm c}\simeq10^{7}\rho_{\infty}$, which corresponds to
\b
 \hat{\phi}_{\infty}=10^{\frac{7}{n+1}}.
\e 
Then, in the case of the smoothed density, the thin-shell condition is given by $ \tilde{m}_{\rm c}^2 L^2 > 10^{7/(n+1)}$.  

\subsection{Numerical analysis}

\begin{figure}[t]
\centering
	\subfigure[profile of $\phi(x)$]{
		\includegraphics[width=7.5cm]{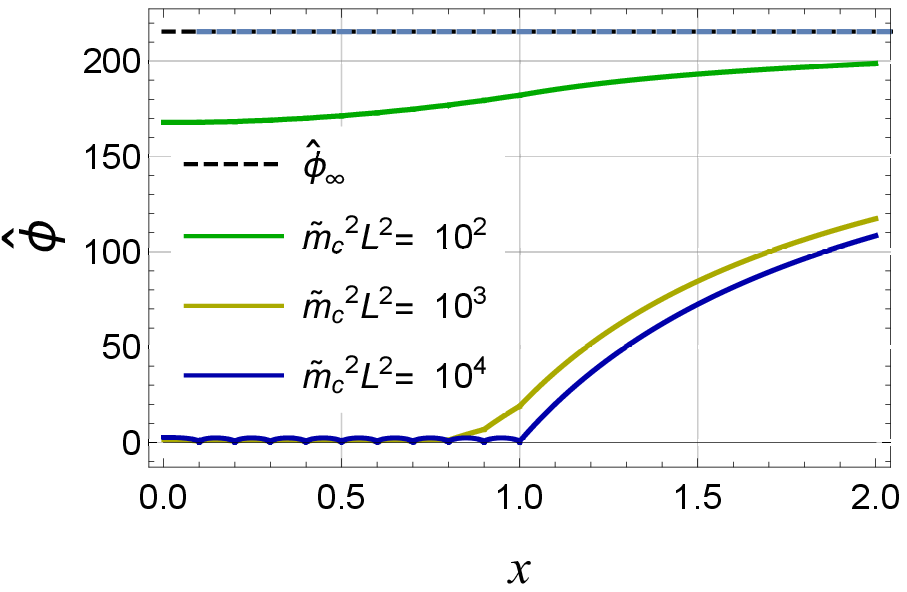}
	}
	\subfigure[$\mathcal R =F_{\phi}/F_{\rm Newton}$]{
		\includegraphics[width=7.5cm]{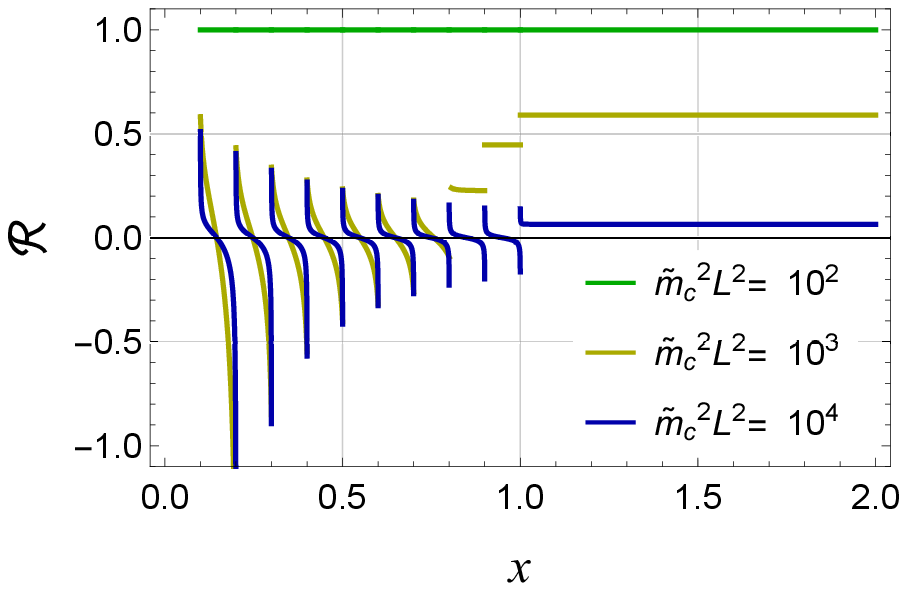}
	}
\caption{These figures show the profile of the chameleon field $\phi$ and the strength of the fifth force divided by the Newtonian gravitational force $\mathcal R$ for $N=10$ and $\tilde{m}_{\rm c}^2L^2=10^2$, $10^3$, $10^4$ respectively.}
\label{v102}
\centering
\end{figure}

First, we consider marginal cases $ \tilde{m}_{\rm c}^2 L^2 \gtrsim 10^{7/(n+1)}$ with a numerical method.  
For simplicity, we choose the power $n$ of the potential as $n=2$ in the analysis.  
Then, the thin-shell condition is given by $\tilde{m}_{\rm c}^2 L^2 \gtrsim 10^{7/3} \sim 200$.
We calculate the field profile by numerically solving the field equations using the shooting method with the junction conditions at each shell as well as the boundary conditions $\dd\hat{\phi}/\dd x|_{x=0} = 0$ and $\lim_{x\rightarrow\infty}\hat
{\phi}(x)=\hat{\phi}_{\infty}$. 
We show the field profile and the value of $\mathcal R$ as functions of $x$ in Fig.~\ref{v102} for $\tilde{m}_{\rm c}^2L^2=10^2, 10^3,$ and $10^4$. 

\begin{figure}[t]
\centering
	\subfigure{
		\includegraphics[width=7cm]{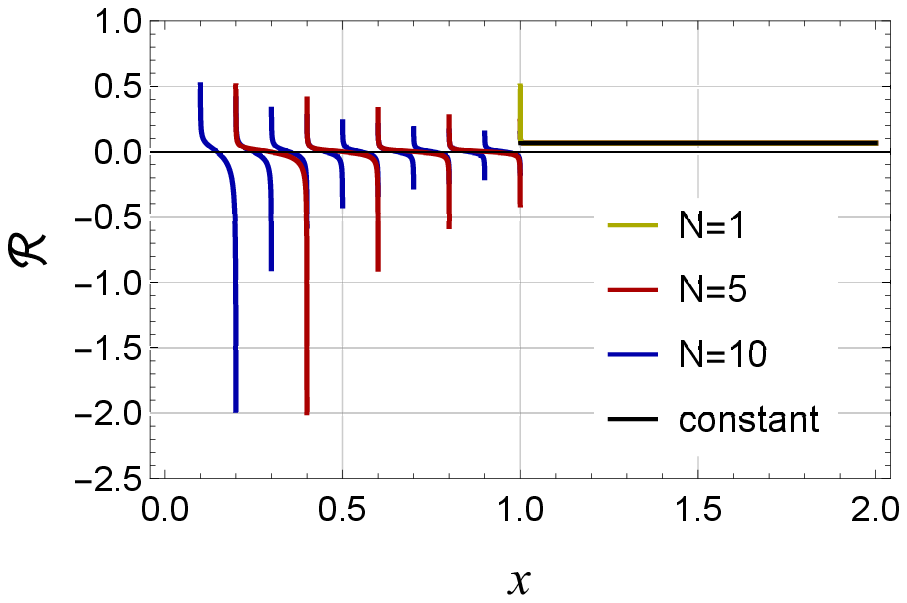}
	}
	\subfigure{
		\includegraphics[width=7cm]{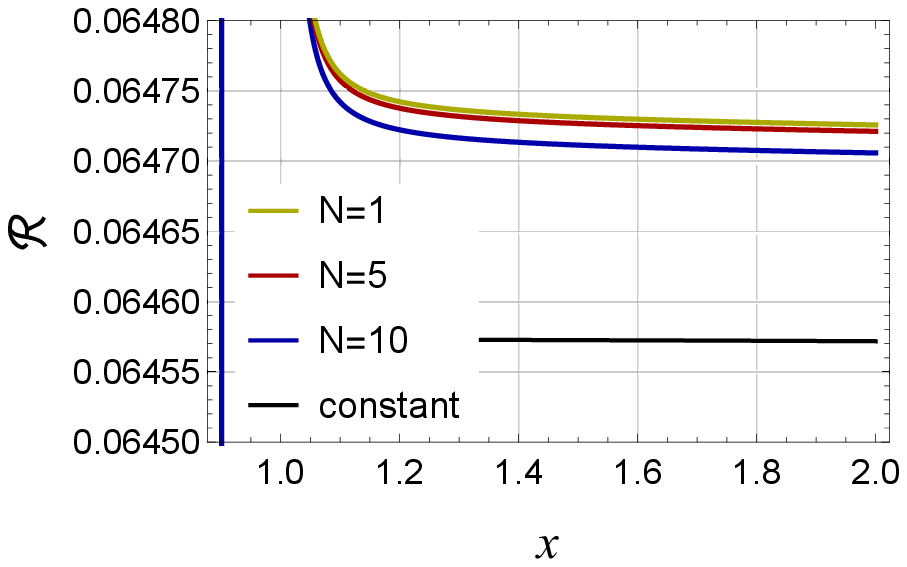}
	}
\caption{This figure shows the variation of the fifth force for different numbers of shells $N = 1, 5,$ and $10$ with $\tilde{m}_{\rm c}^2L^2=10^4$. 
We show an enlarged figure for the outside region in the right panel to 
show the dependence on the number of the shells.  
Each line almost coincides with the one for the smoothed density case in the outside of the outermost shell, $x>1$. 
}
\label{v4nvary}
\centering
\end{figure}

\begin{figure}[t]
	\begin{center}
		\includegraphics[width=8cm]{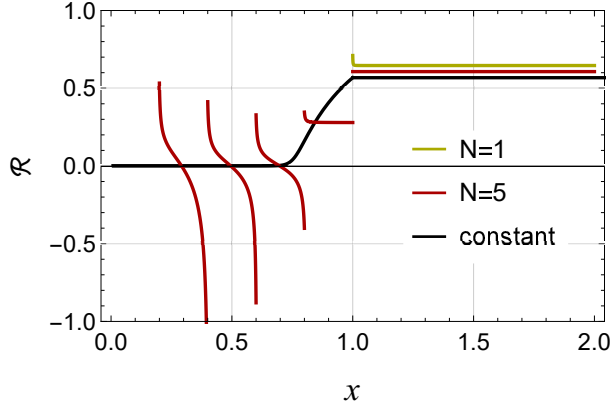}
	\end{center}
\caption{We plot the numerical result of the fifth-force strength divided by the Newtonian 
gravitational force 
for  $\tilde{m}_{\rm c}^2L^2=10^3$ with $N=1$ and $5$. 
The constant density case is also depicted for comparison. }
\label{nvary103}
\end{figure}

In Fig.~\ref{v102}, it is clearly shown that, 
for $\tilde{m}_{\rm c}^2L^2=10^2$, the fifth force is comparable to 
the Newtonian gravitational force everywhere. 
In contrast, for $\tilde{m}_{\rm c}^2L^2=10^4$, the fifth force is 
suppressed compared with the Newtonian gravitational force outside the shell system. 
We also check the dependence on the number of shells $N$. 
In Fig.~\ref{v4nvary}, $\mathcal R$ is depicted as a function of $x$ outside the system for $N = 1, 5,$ and $10$
with $\tilde{m}_{\rm c}^2L^2=10^4$. 
The behavior of $\mathcal R$ is similar to the smoothed-density case 
$\rho = \rho_{\rm c}$ irrespective of the number of shells 
as shown in Fig.~\ref{v4nvary}. 
Therefore, 
the criterion of the thin-shell condition for the screening is applicable to the spherical shell system in the outside region.  
It is worthy  
of note that there is a small but finite deviation even in the outside region.
This finite deviation becomes larger for
the marginal case $\tilde{m}_{\rm c}^2L^2=10^3$
as shown in Fig.~~\ref{nvary103}.
We will discuss it more quantitatively for a large value of $\tilde{m}_{\rm c}^2L^2$ in the subsection \ref{ss:ffout}.
On the other hand, in the inner region, 
the fifth force is not screened well even when the thin-shell condition for the averaged density is satisfied. 
In the subsection \ref{ss:ffin}, we will see that it is true for a larger value of $\tilde{m}_{\rm c}^2L^2$.

\subsection{Analytic approximate solution }

\begin{figure}[t]
	\begin{center}
		\includegraphics[width=7cm]{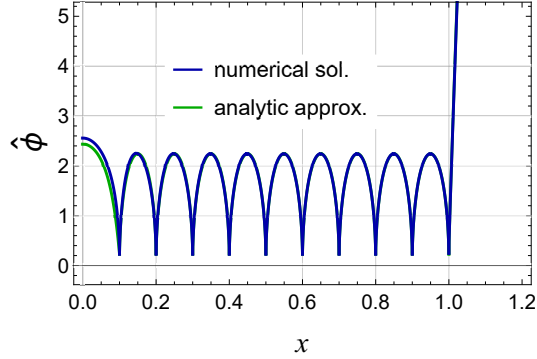}
	\end{center}
\caption{We plot the numerical result and analytical approximate one for $\tilde{m}_{\rm c}^2L^2=10^4$, $N=10$ on the same figure. The blue and green lines correspond 
to the numerical result and 
the analytical approximation, respectively. 
}
\label{v4com}
\end{figure}

The thin-shell condition is well satisfied 
a realistic situation for a galaxy as
$\tilde{m}_{\rm c}^2 L^2 \gtrsim 10^{28-22/(n+1)}$ with $L \sim 10 {\rm kpc}$ \cite{Davis2014}.
Numerical analyses for such a huge value of $\tilde{m}_{\rm c}^2L^2$ 
are very difficult \cite{Kobayashi2008, Upadhye2009}.
Instead of solving the EoM numerically, here, following Ref. \cite{Davis2014}, 
we use an approximation which is valid for a sufficiently large value of $\tilde{m}_{\rm c}^2L^2$. 
We suppose that, for a large value of $\tilde{m}_{\rm c}^2L^2$,
the potential term 
is much larger than the friction term $(2/x)\dd\hat{\phi}/\dd x$
between the shells. 
Then, the EoM can be approximated as follows:
\b
\frac{\dd^2\hat{\phi}}{\dd x^2}+\frac{\tilde{m}_{\rm c}^2L^2}{\hat{\phi}^{n+1}}\simeq0.
\e
The solution for the above equation is given by
\b
\frac{\dd\hat{\phi}}{\dd x}\simeq\pm\sqrt{C+\frac{2\tilde{m}_{\rm c}^2L^2}{n\hat{\phi}^n}},
\label{dphi}
\e
where $C$ is an integration constant.
As is shown in Fig.~\ref{v4com}, the same shape is repeated between the shells. 
The first derivative $\dd\hat{\phi}/\dd x$ vanishes at the middle point
and the profile of $\hat{\phi}$ has a symmetric shape with respect to this middle point.
Assuming a similar repeating structure in the solution for a large value of $\tilde{m}_{\rm c}^2L^2$, 
we can estimate the first derivative at the shell positions 
 as
\b
\left.\frac{\dd\hat{\phi}}{\dd x}\right|_{x=i\cdot\Delta x+0}=\frac{1}{2}\[\frac{\dd\hat{\phi}}{\dd x}\]^+_- \,.
\label{fderiv:shell}
\e
Substituting the junction condition at a shell (\ref{bfdp}) into (\ref{fderiv:shell}), 
we can determine the constant $C$ in terms of the field value at the shell position $\phi_{\rm s}$ as follows:
\b
\tilde{m}_{\rm c}^2L^2\frac{N}{(N+1)(2N+1)}\simeq\sqrt{C+\frac{2\tilde{m}_{\rm c}^2L^2}{n\hat{\phi_{\rm s}}^n}}.
\label{jcshell}
\e
The constant $C$ is written in a simpler form by using the field value at the middle point, $\phi_0$, as $C=-2\tilde{m}_{\rm c}^2L^2/(n\hat{\phi}_0^n)$.
Then, in order for the above approximation to be valid, we need to impose the following condition:
\b
\frac{1}{x}\frac{\dd\hat{\phi}}{\dd x}/\({\frac{\tilde{m}_{\rm c}^2L^2}{\hat{\phi}^{n+1}}}
\simeq\sqrt{\frac{2}{n}}\frac{\hat{\phi}^{n/2+1}}{\tilde{m}_{\rm c}Lx} \sqrt{1-\({\hat{\phi}/\hat{\phi}_0}^n}
\ll1. 
\label{appvaconv}
\e
Our numerical results in  Fig. \ref{v102} show that the field value $\hat{\phi}$ varies at most by $\Delta \hat{\phi}/\hat{\phi} = {\cal O}(1)$ in the inner regions. 
Therefore, our approximation is valid for a sufficiently large value of $\tilde{m}_{\rm c}^2 L^2$.
 
The approximation \eqref{appvaconv} cannot be applied to the region near the center $\tilde{m}_{\rm c}Lx \ll 1$ and then neither the solution (\ref{dphi}).
In this region, we use the following asymptotic expansion of $\hat{\phi}$ 
inside the innermost shell: 
\b
\hat{\phi}=c_0-\frac{1}{6}c_0^{-n-1}\tilde{m}_{\rm c}^2L^2x^2-\frac{1}{120}(n+1)c_0^{-2n-3}\tilde{m}_{\rm c}^4L^4x^4+\cdots,
\label{center}
\e
where $c_{0}$ is the field value at the origin.
This expansion is valid for sufficiently small $\tilde{m}_{\rm c}Lx$.
We can construct the field profile by jointing the 
approximate solutions (\ref{center}), (\ref{dphi}) and (\ref{phio}) 
at each shell with the junction condition (\ref{bfp}). 
In Fig.~\ref{v4com}, we show that the analytic approximation 
agrees well with 
the numerical result for $\tilde{m}_{\rm c}^2L^2=10^4$ and $N=10$. 
The deviation between the analytic approximation and the numerical result is less than several percents.


\subsection{Fifth Force outside the System}\label{ss:ffout}

Let us evaluate the value of $\mathcal R$ 
in the outside of the system
by using the analytic approximation given  
in the previous subsection. 
For simplicity, we concentrate on a specific form of the potential with $n=2$, then 
Eq.~(\ref{dphi}) can be easily solved as
\b
\hat{\phi}=\sqrt{C_1-\frac{\tilde{m}_{\rm c}^2L^2}{C_1}(x+C_2)^2},
\e
where $C_1$ and $C_2$ are integration constants. 
The integration constants can be rewritten by 
using the field value at a shell $\hat{\phi}_s$ as
\b
\hat{\phi}(x)=\sqrt{\frac{\hat{\phi}_{\rm s}^2+\sqrt{\hat{\phi}_{\rm s}^4+\tilde{m}_{\rm c}^2L^2/N^2}}{2}-\frac{2\tilde{m}_{\rm c}^2L^2(x-x_0)^2}{\hat{\phi}_{\rm s}^2+\sqrt{\hat{\phi}_{\rm s}^4+\tilde{m}_{\rm c}^2L^2/N^2}}} \,,
\label{phiap}
\e
where $x_0$ is the value of $x$ at the middle of the interval and we have assumed $\dd \hat{\phi}/\dd x|_{x=x_0}=0$ ,which is suggested from the numerical calculation. 
According to the junction condition (\ref{jcshell}), we can determine the field value at the shell position $\hat{\phi}_{\rm s}$ by the following equation: 
\b
\hat{\phi}_{\rm s}\left(\hat{\phi}_{\rm s}^2+\sqrt{\hat{\phi}_{\rm s}^4+\tilde{m}_{\rm c}^2L^2/N^2}\right)=\frac{(N+1)(2N+1)}{N^2}. 
\label{eqofphis}
\e
From the above equation, we obtain the following behavior depending on the value of the parameter $m_{\rm c}L/N=\Delta x/\lambda_{\phi}$:
\b
\hat{\phi}_{\rm s}\sim
\left\{
	\begin{array}{cc}
		1&(\tilde{m}_{\rm c}^2L^2/N^2\ll1)\\
		2N/(\tilde{m}_{\rm c}L)&(\tilde{m}_{\rm c}^2L^2/N^2\gg1)
	\end{array}
\right.
\label{phislimit}
\e
with estimating the right-hand side of Eq. \eqref{eqofphis} to be ${\cal O}(1)$. 
Therefore, it is assured that $\hat{\phi}_{\rm s}$ is less than ${\cal O}(1)$.
If the approximation 
is valid even at the outermost shell,  
the field value at the outermost shell is also given by 
$\hat{\phi}_{\rm s}$.
Then, we can estimate the fifth force outside the object from the equation (\ref{phio}) as 
\b
\lim_{m_\infty L\rightarrow0}F_{\phi}=\frac{\beta \phi_{\rm c}}{M_{\rm pl}L}(\hat{\phi}_{\infty}-\hat{\phi}_{\rm s})\frac{1}{x^2} \,,
\label{ffap}
\e
in the limit $m_{\infty}L\rightarrow 0$.
The effect of the inhomogeneity
on the fifth force outside the object
can be calculated by taking the difference between  
Eqs.~(\ref{ffap}) and 
(\ref{fiff_uni}) as follows: 
\b
\lim_{m_{\infty}L\rightarrow0}\left(F_{\phi}-F_{\phi}^{\rm con}\right)
=\frac{\beta \phi_{\rm c}}{M_{\rm pl}L}(1-\hat{\phi}_{\rm s})\frac{1}{x^2}. 
\label{outf}
\e
We see that, from Eq.~(\ref{phislimit}), the value of $\phi_{\rm s}$ approaches to unity 
and thus $F_\phi \to F_\phi^{\rm con}$ for $\tilde{m}_{\rm c}^2L^2/N^2 = (\Delta x/\sqrt{3}\lambda_{\phi})^2 \rightarrow 0$.
It is also noteworthy that 
the difference between $F_\phi$ and $F_\phi^{\rm con}$ is suppressed by the 
factor $1/(\tilde{m}_{\rm c}^2L^2)$ compared to the Newtonian gravitational force as follows:
\b
\lim_{m_{\infty}\rightarrow0}\left|\frac{F_{\phi}-F_{\phi}^{\rm con}}{F_{\rm Newton}}\right|=\frac{6\beta^2}{\tilde{m}_{\rm c}^2L^2}(1-\hat{\phi}_{\rm s}). 
\label{diffef}
\e
The above expression is valid only for a large value of $\tilde{m}_{\rm c}^2L^2$ but suggests that 
the difference between $F_\phi$ and $F_\phi^{\rm con}$ may be non-negligible for marginal cases such as $\tilde{m}^2_cL^2=10^3$. 


\subsection{Fifth Force inside the System}\label{ss:ffin}

\begin{figure}[t]
	\begin{center}
		\includegraphics[width=7cm]{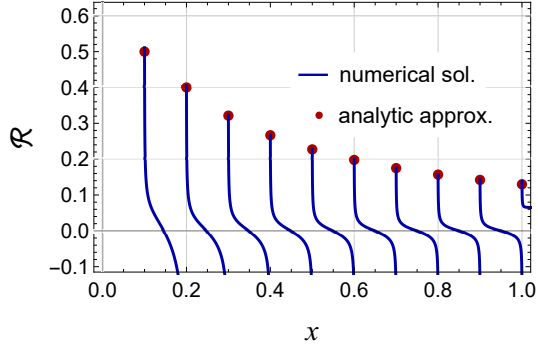}
	\end{center}
\caption{
The value of $\mathcal R$ is depicted as a function of $x$. 
The spiky blue lines show the result of numerical integration for $N=10$ and $\tilde{m}_{\rm c}^2L^2=10^4$. 
The red points show the analytic approximation at the position of each shell given by Eq. \eqref{inf} with substituting $i=x/N$. 
}
\label{ffanvsnu}
\end{figure}

As we have already mentioned in Sec.~\ref{SinSSS}.B, the field profile is approximately symmetric at each shell, 
so that the derivative of the field has the same absolute value but  
the opposite sign at each side. 
Then, the value of $\mathcal R$ at each shell can be straightforwardly evaluated by the 
junction condition \eqref{fderiv:shell} and the form of Newtonian gravitational force \eqref{newton} as follows: 
\b
\mathcal R=\frac{6\beta^2i}{(i+1)(2i+1)}.
\label{inf}
\e
The maximum value $\mathcal R_{\rm max} = \beta^2$ is realized at the innermost shell for $i=1$ irrespective of a value of $\tilde{m}_{\rm c}^2 L^2$. 
It is to be noted that the value (\ref{inf}) is obtained without specifying the potential form.
We can confirm the validity of the approximation \eqref{inf} by comparing it with the numerical result~(see Fig.~\ref{ffanvsnu}).
This result is very suggestive in the following sense: 
even if the Compton wavelength is sufficiently smaller than the size of the object, 
so that the fifth force is screened outside the object, 
the value of the fifth force can be comparable to the Newtonian gravitational force 
in the shell system.


\section{Effect of the finite width and the origin of the fifth force enhancement in our model}
\label{thick}

In this section, we discuss how the fifth force appearing 
 in the previous section depends on the width of the shells.

\subsection{Fifth Force outside the System}

\begin{figure}[htbp]
\begin{center}
\includegraphics[clip,width=8cm]{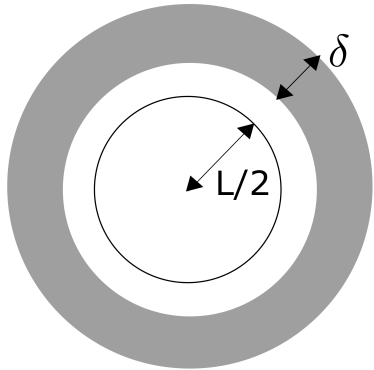}
\caption{A schematic figure of the thick outer shell and thin inner shell system.}
\label{sss2}
\end{center}
\end{figure}

\begin{figure}[t]
\centering
	\subfigure{
		\includegraphics[width=6.4cm]{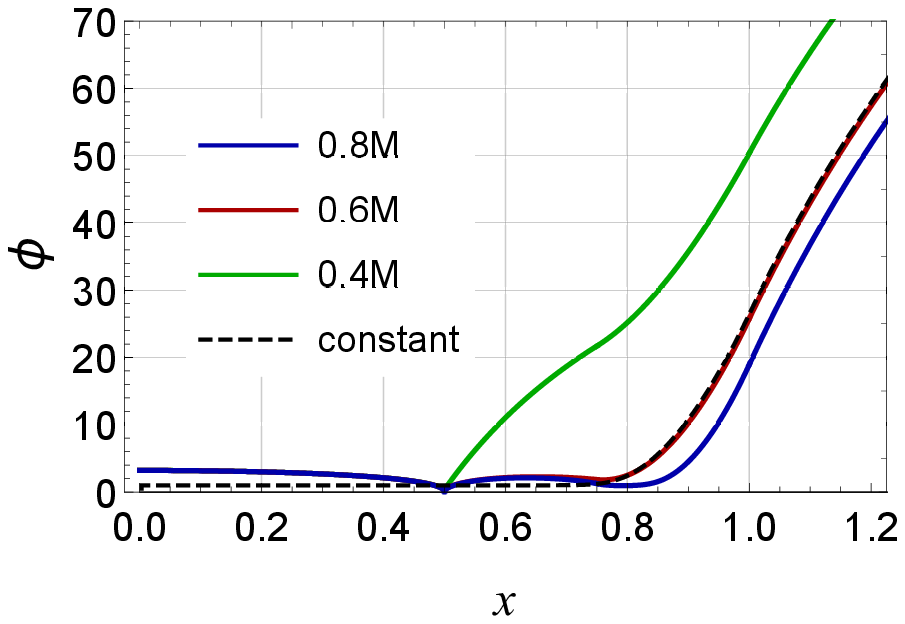}
	}
	\subfigure{
		\includegraphics[width=8.6cm]{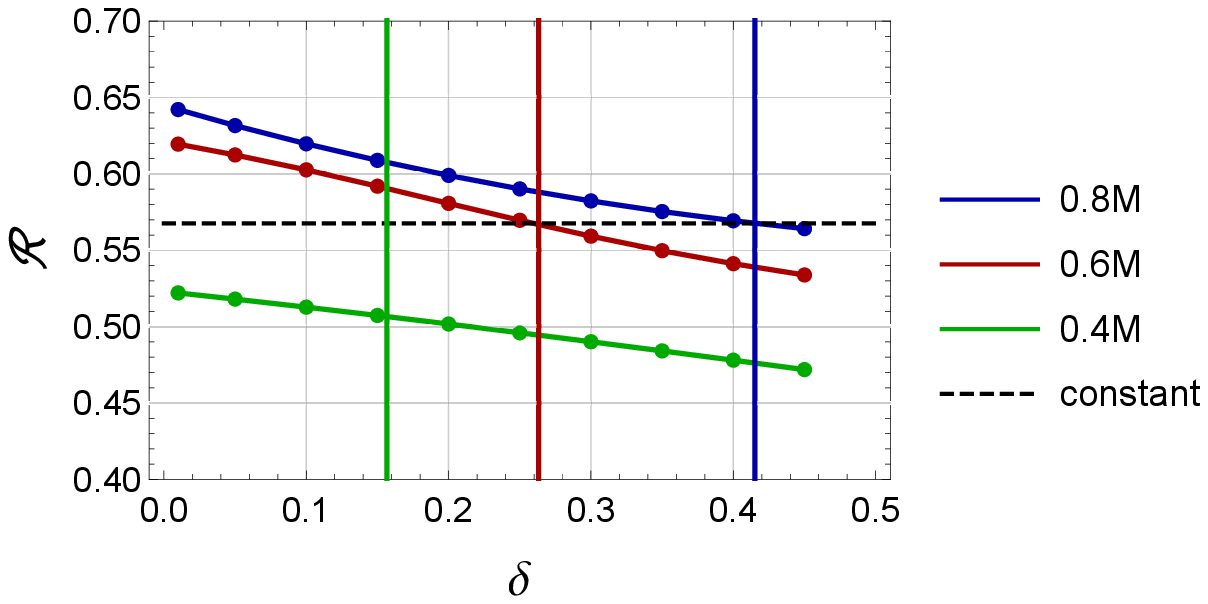}
	}
\caption{
The field profile and the value of $\mathcal{R}=F_{\phi}/F_{\rm Newton}$ for $M_{\rm o}=0.8M, 0.6M$ and $0.4M$ with $\tilde{m}_{\rm c}^2L^2=10^3$ is depicted.
The field profile is shown with $\delta=0.25$ in the left panel and $\mathcal{R}$ is calculated at the same point, $x=2$,  as a function of $\delta$ in the right panel. 
The value of $\mathcal R$ for the uniform density case is shown by the dashed line. The vertical lines on the right panel correspond to the position of $\delta_c$ respectively.
}
\label{outsideR}
\centering
\end{figure}

First, we examine the fifth force outside the object with changing the thickness $\delta$ of the outermost shell. 
For simplicity, we divide the total mass of the system into the outermost thick shell and 
the other inner thin shell at the radius $L/2$(see Fig.~\ref{sss2}). 
The value of $\mathcal{R}=F_{\phi}/F_{\rm Newton}$ at the outer surface of the thick shell is depicted as a function of $\delta$ for each value of the outer shell mass $M_{\rm o}$ in the right panel of Fig.~\ref{outsideR} for $\tilde{m}_{\rm c}^2L^2=10^3$. 
The field profile is also shown on the left panel for the same parameters with $\delta=0.25$ .
We note that, in this setup, $1-x_{roll}$ in Sec. III is estimated to be about $0.2$, which gives a typical length for the scalar field to settle down to the minimum. 
Let us define $\delta_{\rm c}$ as the value of $\delta$ for which the density of the thick shell is equal to that of 
the uniform density case. 
The vertical line shows the value of $\delta_{\rm c}$ for each value of $M_{\rm o}$. 
As is shown in Fig.\ref{outsideR}, we obtain a smaller fifth force value for a thicker shell. 
For $M_{\rm o}=0.6M$ and $0.8M$ cases, the lines intersect with the dashed line of the uniform density 
when the density of the thick shell is equal to that of the uniform density case. 
As can be easily seen from the field profile shown in the left panel of  Fig.~\ref{outsideR}, in these cases, the scalar field is settle down to the minimum of the effective potential for the constant density and gives the same field profile in the outer region.
While, for $M_{\rm o}=0.4M$ case, we find $\delta_{\rm c}<1-x_{\rm roll}$, and 
the value of $\mathcal R$ dose not exceed that for the uniform density case. 
In summary, 
the enhancement of the fifth force outside the system is due to the higher density near the surface. 
Therefore, we may conclude that 
the fifth force outside the system can be significantly different from the uniform density case 
only if the density distribution within the damping depth 
is significantly different.

\subsection{Fifth Force inside the System}

\begin{figure}[t]
\centering
	\subfigure[profile of $\phi(x)$]{
		\includegraphics[width=7.5cm]{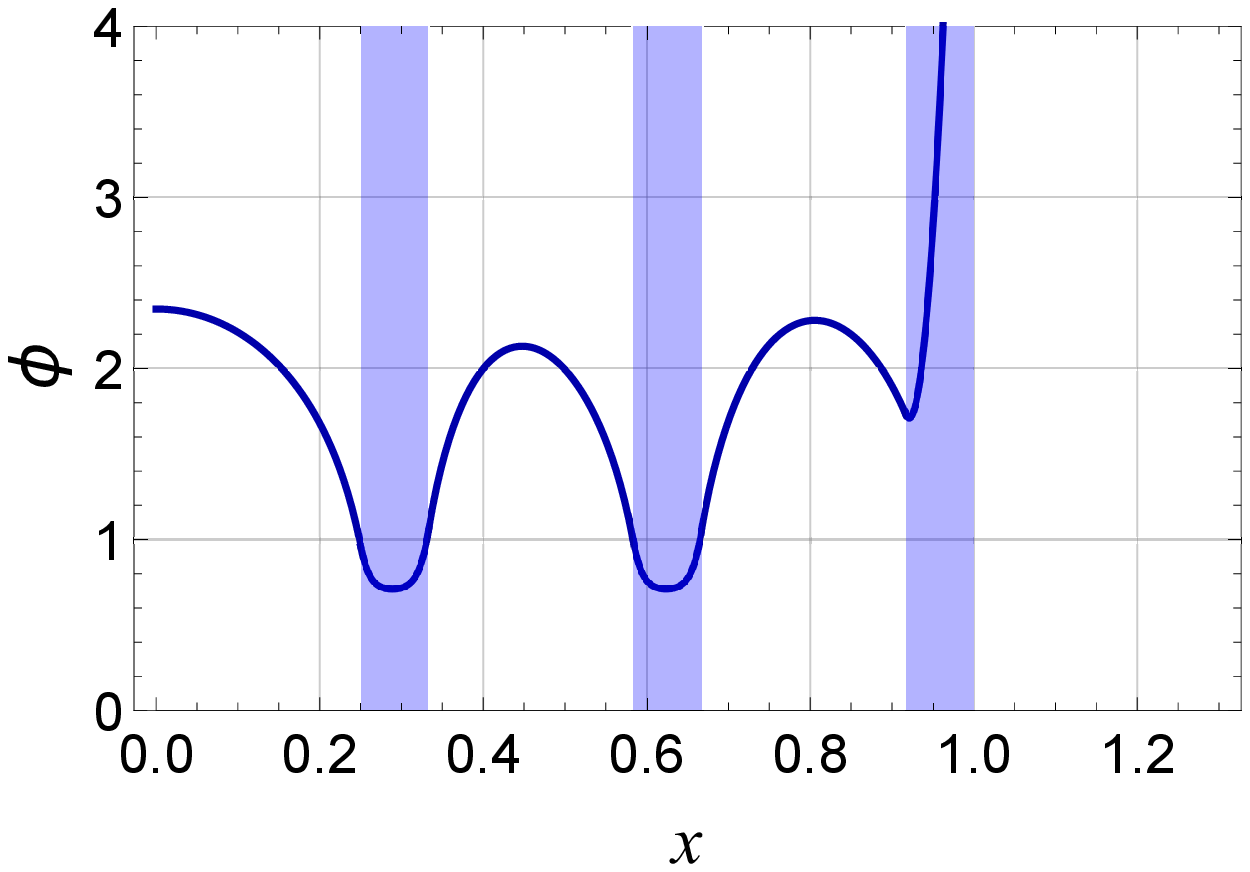}
	}
	\subfigure[$\mathcal R =F_{\phi}/F_{\rm Newton}$]{
		\includegraphics[width=7.5cm]{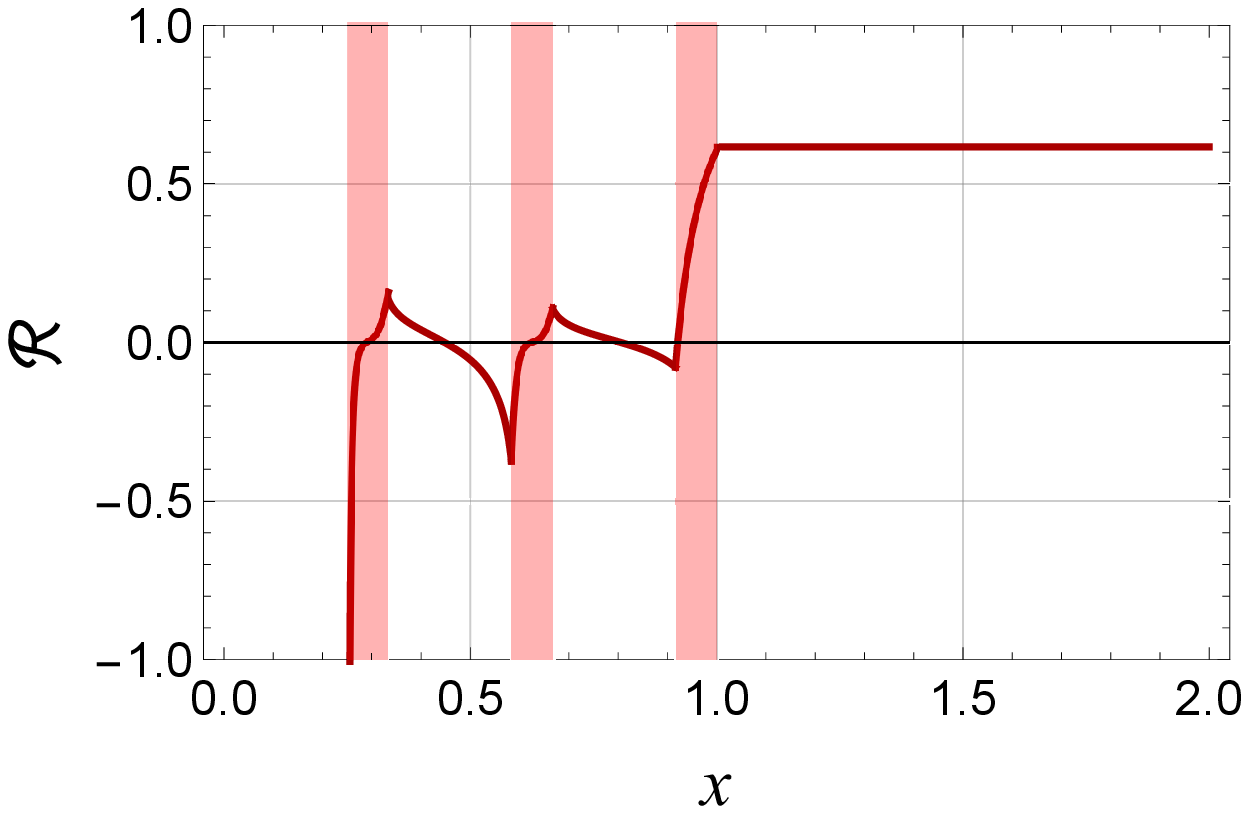}
	}
\caption{These figures show the profile of the chameleon field $\phi$ and the strength of the fifth force divided by the Newtonian gravitational force $\mathcal R$ for $N=3$, $\tilde{m}_{\rm c}^2L^2=10^3$ and $a=1/12$. 
The blue and red regions represent the shell regions. }
\label{thick_numeric}
\centering
\end{figure}

\begin{figure}[t]
	\begin{center}
		\includegraphics[width=8cm]{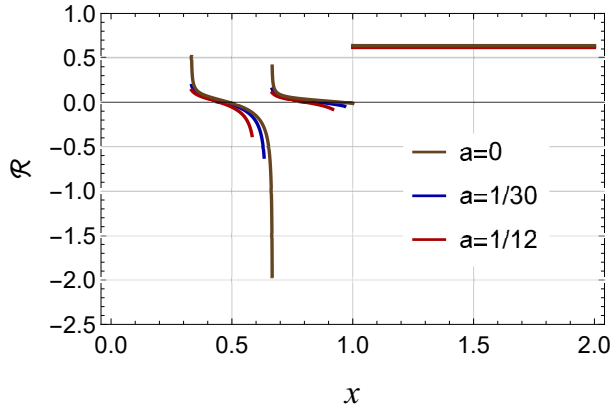}
	\end{center}
\caption{We plot the ratio $\mathcal{R}=F_{\phi}/F_{\rm Newton}$ for $a=0, 1/12, 1/30$
with $\tilde{m}_{\rm c}^2L^2=10^3$, $N=3$. 
The profiles are plotted only in the vacuum regions to make the difference easier to see. }
\label{avary}
\end{figure}

Next, we consider the effect of the thickness of the shell on the inside part.  
For simplicity, we assume each shell has the identical
width and density. The shells regularly foliate the spherical region with a fixed interval.
It is also assumed that the size and total mass of the system is given by $L$ and $M_{\rm tot}$ respectively as before.
Then, the smoothed density $\rho_c$ is also fixed. 
Denoting the width of the shells by $aL$, which should be less than $L/N$, the density of each shell is given by
\begin{align}
\rho&=\rho_c\frac{V_{\rm s}}{V_{\rm shell}}\n\\
&=\rho_c\frac{2N}{(N+1)(2N+1)a-3N(N+1)a^2+2N^2a^3} \sim \frac{\rho_c}{Na} \,,
\label{shellrho}
\end{align}
where $V_{\rm s}$ is the volume of the whole system and $V_{\rm shell}$ is the total volume of all shells. 
The last equality is satisfied for large $N$ and small $a$.
We calculate the field profile with the same boundary conditions as those in the previous calculations.
The result has not changed qualitatively from the previous ones as shown in Fig.~\ref{thick_numeric}. 
The large fifth force appears at each shell although its amplitude becomes smaller.
From Fig.~\ref{avary}, we can see the profile approaches to that for the infinitely thin shells as the width $aL$ decreases. 

Let us consider the origin of the large fifth force for thin shell cases. 
From Eq. (\ref{def:mass}), we can find that the effective mass in the shell is enhanced by the factor $(Na)^{-\frac{1}{2}\frac{n+2}{n+1}}$ compared with that for the uniform-density object.
On the other hand, 
the length scale of a shell $L_{\rm shell}$ along the radial direction is given by its width $aL$.
Then, we can roughly estimate the screening parameter for each shell as
\b
(\tilde m_{\rm eff} L_{\rm shell})^2 \sim N^{-\frac{n+2}{n+1}} a^{\frac{n}{n+1}} (\tilde m_{\rm c} L)^2 \,.
\label{thin_pm_shell}
\e
Therefore, when the number of the shells $N$, or the mass of each shell, is fixed, 
the shell becomes  totally unscreened for the 
limit $a \to 0$. 
This is the origin of the large fifth force in the inner region. 
This argument indicates that, even when the chameleon screening mechanism is working for a total system,  
the components of the system can be unscreened and then a large fifth force can appear in its inside, depending on the shape of the components.


\section{Conclusion}
\label{conclusion}

We investigated the chameleon screening mechanism for inhomogeneous density profiles. 
For some specific density profile with high density contrasts, 
it is expected that the chameleon field cannot trace the minimum of the 
varying potential and the smoothing of the density may not be justified.
To explicitly show it, we considered one of the simplest examples, the spherical shell system composed of 
a set of concentric shells, where there is no potential minimum at 
any radius and the chameleon field cannot be stable 
by a large mass as usually assumed for a successful screening of the fifth force.

The results show 
that the fifth force can be screened outside the system if the so-called thin-shell 
condition is satisfied for the smoothed average density as in the case of a constant density profile. 
The screening mechanism successfully works for a cluster of unscreened objects 
if the cluster satisfies the thin-shell condition on average.
However, we find the inhomogeneity near the surface can contribute to the fifth force value for the marginal screening case. 

The field profile inside the system can be significantly different 
from the smoothed density case for the shell system. 
We derived an analytic approximate expression for the fifth force inside the system 
with the help of insights from the numerical results. 
In our simple toy model, irrespective of the other model parameters, 
the maximum value of the ratio between the fifth force and 
the Newtonian gravitational force is given by $\beta^2$ with $\beta$ being the dimensionless coupling constant for 
the conformal coupling between the standard matter and the chameleon field. 
Since the value of $\beta$ is usually assumed to be in the order of 1, 
our result suggests the possibility that the fifth force can be significantly large inside an object with a highly inhomogeneous density profile. 
Due to the fact that this result is irrelevant to the property of the effective Compton wavelength, 
the same concern may exist in other fifth force models which 
have a circumstance dependent screening mechanism, such as the 
symmetron \cite{Hint2010} 
and the environmentally dependent dilaton \cite{Brax2011}. 
One should not feel complacent about the wellbehavedness of the fifth-force field 
with an averaged density distribution. 
A significant fifth force strength can be induced inside an inhomogeneous object 
depending on the shape of inhomogeneity. 
Since it does not follow the inverse square law, unlike the case of Newtonian gravitational force, the configuration of outer shell affects the fifth force inside it. 
As shown in 
Fig.~\ref{v102}, the fifth force works in the direction of collecting matters to each shell.
Then,  it may cause new instability other than the one caused by the usual gravitational attraction 
and should be investigated more carefully as a factor that may affect the structure formation.

\acknowledgements
We would like to thank Prof. Shin'ichi Nojiri for his useful comments.
We also would like to thank the anonymous referee for helpful suggestions. 
This work was supported by JSPS
KAKENHI Grant Numbers JP16K17688, JP16H01097 (CY) and JP17K14286 (RS).

\bibliographystyle{apsrev4-1}

\end{document}